\documentclass[runningheads]{llncs}
\usepackage[T1]{fontenc}
\usepackage{amsfonts, amssymb, amsmath}
\usepackage{multirow,booktabs}
\usepackage{subcaption}
\usepackage{algorithm}
\usepackage{algpseudocode}

\usepackage{tikz}
\usetikzlibrary{shapes.geometric, arrows.meta, positioning, calc, backgrounds, fit}
\usepackage{graphicx}
\begin{document}
\title{Counterexample Guided Branching via Directional Relaxation Analysis in Complete Neural Network Verification}

\author{Jingyang Li\inst{1} \and Fu Song\inst{2}\and Guoqiang Li\inst{1}}

\institute{Shanghai Jiao Tong University, Shanghai 200240, China\\
\email{\{lijjjjjj, li.g\}@sjtu.edu.cn}
\and
Institute of Software, Chinese Academy of Sciences, Beijing 100190, China\\
\email{songfu@ios.ac.cn}
}

\maketitle              % typeset the header of the contribution
\begin{abstract}
Deep Neural Networks demonstrate exceptional performance but remain vulnerable to adversarial perturbations, necessitating formal verification for safety-critical deployment. To address the computational complexity of this task, researchers often employ abstraction-refinement techniques that iteratively tighten an over-approximated model. While structural methods utilize Counterexample-Guided Abstraction Refinement, state-of-the-art dataflow verifiers typically rely on Branch-and-Bound to refine numerical convex relaxations. However, current dataflow approaches operate with blind refinement processes that rely on static heuristics and fail to leverage specific diagnostic information from verification failures. In this work, we argue that Branch-and-Bound should be reformulated as a Dataflow CEGAR loop where the spurious counterexample serves as a precise witness to local abstraction errors. We propose DRG-BaB, a framework that introduces the Directional Relaxation Gap heuristic to prioritize branching on neurons actively contributing to falsification in the abstract domain. By deriving a closed-form spurious counterexample directly from linear bounds, our method transforms generic search into targeted refinement. Experiments on high-dimensional benchmarks demonstrate that this approach significantly reduces search tree size and verification time compared to established baselines.
% The scalability of Branch-and-Bound (BaB) algorithms for complete neural network verification is fundamentally bottlenecked by the efficacy of branching heuristics. Existing strategies typically rely on static geometric properties or average-case sensitivity, which often fail to pinpoint the specific relaxation responsible for proof failures in the abstract domain. In this paper, we propose \textsc{DRG-BaB}, a novel framework that reformulates the branching process as a Counterexample Guided Abstraction Refinement (CEGAR) loop. Our key insight is that the spurious counterexample encountered during failed verification is not merely a byproduct of relaxation, but a precise \textit{witness} to the abstraction error. Based on this, we introduce the \textit{Directional Relaxation Gap} (DRG) heuristic. DRG identifies optimal branching neurons by quantifying the relaxation error at the witness point, specifically conditioned on the gradient direction of the safety loss to ensure global relevance. We integrate our framework into state-of-the-art verifiers, including both LP-based tools and GPU-accelerated dual solvers like $\alpha$-$\beta$-CROWN. Experimental results on high-dimensional benchmarks including MNIST and CIfAR-10 demonstrate that \textsc{DRG-BaB} significantly reduces the search tree size and computational overhead compared to established geometric and gradient-based baselines, showcasing the power of counterexample-guided refinement in neural network verification.

\end{abstract}

\section{Introduction}
Deep Neural Networks (DNNs) have demonstrated unparalleled performance in perception and control tasks. However, their deployment in safety-critical systems (e.g., autonomous driving and avionics) remains cautious due to their vulnerability to adversarial perturbations~\cite{szegedy2013intriguing,goodfellow2014explaining}, where imperceptible input changes can lead to catastrophic failures. Consequently, \textit{Formal Verification}, which provides mathematical guarantees of network properties, has emerged as a fundamental requirement for trustworthy AI~\cite{huang2017safety}.

The complete verification of ReLU networks is known to be NP-complete~\cite{katz2017reluplex}. To address this complexity, the principle of abstraction-refinement~\cite{cousot1977abstract} is generally adopted. By reasoning over an over-approximated (abstract) model of the network, verifiers can efficiently prove safety. However, due to information loss, the abstraction may fail to prove a valid property. To recover completeness, the verifier must iteratively refien the abstract model to eliminate precision erros. In the landscape of neural network verification, two distinct paradigms are observed: structural and dataflow abstraction-refinement.

Sturctural abstraction-refinement is exemplified by methods that simplify the network topology (e.g., neuron clustering). Such paradigm often follows the classif Counterexample-Guided Abstraction Refinement (CEGAR) framework~\cite{clarke2000counterexample} that uses the spurious counterexample to refine specific parts of the abstract structure. As for dataflow abstraction-refinement, it is widely adopted in state-of-the-art complete verifiers~\cite{wang2021beta,wang2018neurify}. The abstraction is numerical rather than structural, using convex relaxations~\cite{singh2019deep,zhang2018efficient} to bound intermediate values. Refinement is achieved via Branch-and-Bound (BaB), which splits input or activation domains to tighten the relaxation.

While structural methods explicitly leverage CEGAR, the dataflow paradigm operates with a blind refinement process. Existing branching heuristics refine either on static geometric properties (e.g., interval width) or average-case sensitivity (e.g., BaBSR~\cite{bunel2020branch}). These heuristcs fail to adequately utilize the diagnostic information from the verification failure in the abstract domain, that is, the specific state of the relaxed network when it violates the property.

\textbf{Our Insight:} We argue that efficient BaB should be reformulated as a Dataflow CEGAR loop. When a relaxation-based verifier fails to prove a property, the underlying solver either implicitly finds a specific point $x^*$ where the relaxed network violates the specification. $x^*$ is either a true counterexample that falsify the property on the concrete network, or a spurious counterexample to the concrete network. The latter is not merely a byproduct of failure, but a precise witness to the abstraction error. $x^*$ pinpoints exactly where the convex relaxation is too loose relative to the safety specification.

Based on this insight, we propose DRG-BaB, a novel framework that integrates CEGAR-sytle guidance into the BaB process. Instead of branching on neurons that are merely "wide" or "sensitive", we introduce the Directional Relaxation Gap (DRG) heuristic. DRG analyzes the spurious counterexample to identify the specific neurons whose relaxation errors activaly contribute to the falsification in abstract domain. By prioritizing these neurons for branching, we transform the refinement step from a generic search into a targeted correction of the abstraction, significantly pruning the search tree.

\begin{itemize}
    \item \textbf{A Dataflow CEGAR Framework for Neural Network Verification. } We formalize the Branch-and-Bound search as a Counterexample-Guided Abstraction Refinement loop. Unlike existing approaches that treat verification failure as a generic signal to branch, we interpret the spurious counterexample as a precise witness to local abstraction errors. This paradigm shift allows us to transform the branching strategy from a static heuristic into a dynamic, goal-driven refinement process.

    \item \textbf{Directional Relaxation Gap Heuristic.} To address the "attribution problem" of identifying which neurons are responsible for the spurious witness, we propose the DRG metric. By quantifying the relaxation error conditioned on the gradient direction of the safety loss, DRG effectively distinguishes active errors (which enable the violation) from inactive ones (which are geometrically loose but harmless). This resolves the "Wide but Tight" dilemma that misleads traditional width-based heuristics.

    \item \textbf{Generalization and Performance.} We integrate our framework into the state-of-the-art GPU-accelerated verifier, $\alpha$-$\beta$-CROWN. Crucially, we derive a closed-form construction of spurious counterexample directly from the linear bounds (CROWN), eliminating the need for computationally expensive exact slolution. Extensive experiments on high-dimensional benchmarks (CIfAR-10 and MNIST) demonstrate that DRG-BaB reduces the search tree size and verification time compared to established baselines, confirming the efficacy of witness-guided refinement in large-scale verification.
\end{itemize}

\section{Preliminaries}
\label{sec:preliminaries}
In this section, we formalize the neural network verification problem and provide the necessary background on linear relaxation techniques and the Branch-and-Bound (BaB) framework.

\subsection{Neural Networks and Verification Problem}

We consider an $L$-layer feed-forward neural network $f: \mathbb{R}^{n_0} \to \mathbb{R}^{n_L}$. Let $z^{(i)}$ denote the pre-activation vector and $\hat{z}^{(i)}$ denote the post-activation vector at layer $i$. The propagation is defined as:
\begin{equation}
    z^{(i)} = W^{(i)} \hat{z}^{(i-1)} + b^{(i)}, \quad \hat{z}^{(i)} = \sigma(z^{(i)}), \quad i=1,\dots,L
\end{equation}
where $W^{(i)}$ and $b^{(i)}$ are weights and biases, $\hat{z}^{(0)} = x$ is the input, and $\sigma(\cdot)$ is the element-wise activation function. In this work, we focus on the Rectified Linear Unit (ReLU), i.e., $\sigma(y) = \max(0, y)$.

Given an initial input domain $\mathcal{D}_0 \subseteq \mathbb{R}^{n_0}$ (typically a hyper-rectangle defined by $x_L \leq x \leq x_U$), the verification problem asks whether the output $f(x)$ satisfies a safety property $\mathcal{P}$ for all $x \in \mathcal{D}_0$. Without loss of generality, we consider robustness properties defined by a set of linear constraints on the output logits. This is equivalent to verifying that the \textit{margin function} $m(x)$ is strictly positive:
\begin{equation}
    \forall x \in \mathcal{D}_0, \quad m(x) := C f(x) > 0
\end{equation}
where $C$ is a specification matrix. If the global lower bound $\min_{x \in \mathcal{D}_0} m(x) > 0$, the property holds (Safe). If there exists a concrete input $x^* \in \mathcal{D}_0$ such that $m(x^*) \leq 0$, then $x^*$ is a valid counterexample (Unsafe).

\subsection{Linear Relaxation and Branch-and-Bound Verification}
\label{subsec:rel_and_bab}

To circumvent the NP-completeness of exact reachability~\cite{katz2017reluplex}, state-of-the-art verifiers employ \textbf{Linear Relaxation} to over-approximate network behavior for efficient bound propagation~\cite{zhang2018efficient,singh2019deep}.
Consider a neuron $j$ in layer $i$ with pre-activation bounds $[l_j^{(i)}, u_j^{(i)}]$. If the neuron is \textbf{unstable} (i.e., $l_j^{(i)} < 0 < u_j^{(i)}$), the non-linear constraint $\hat{z} = \text{ReLU}(z)$ is relaxed into a convex region defined by a linear upper bound connecting $(l_j^{(i)}, 0)$ and $(u_j^{(i)}, u_j^{(i)})$, and a linear lower bound $\hat{z}_j^{(i)} \geq \alpha_j^{(i)} z_j^{(i)}$, where $\alpha_j^{(i)} \in [0, 1]$ is an optimizable parameter~\cite{xu2021fast}.
The geometric area between these linear approximations and the exact ReLU function constitutes the \textbf{relaxation error}.
By propagating these constraints, a valid symbolic lower bound $\underline{m}(x)$ over the input domain $\mathcal{D}_0$ is derived.

However, due to the inherent relaxation error, $\underline{m}(x)$ may be too loose to prove safety (i.e., $\min_{x \in \mathcal{D}_0} \underline{m}(x) < 0$). To recover completeness, the \textbf{Branch-and-Bound (BaB)} framework recursively partitions the search space to refine the abstraction.
Rather than splitting the input space $\mathcal{D}_0$, which suffers from the curse of dimensionality, we focus on branching on intermediate neurons.
Formally, a verification sub-problem is defined over a sub-domain $\mathcal{D}$ augmented with a set of \textbf{split constraints} $\mathcal{S} = \{(i, j, s) \mid s \in \{+1, -1\}\}$, where $s=+1$ enforces an active state ($z_j^{(i)} \ge 0$) and $s=-1$ enforces an inactive state ($z_j^{(i)} < 0$). The sub-domain represents:
\begin{equation}
    \mathcal{D} = \mathcal{D}_0 \cap \bigcap_{(i, j, +1) \in \mathcal{S}} \{x \mid z_j^{(i)}(x) \geq 0\} \cap \bigcap_{(i, j, -1) \in \mathcal{S}} \{x \mid z_j^{(i)}(x) < 0\}
\end{equation}
Within $\mathcal{D}$, the neurons in $\mathcal{S}$ become linear, thereby eliminating their individual relaxation errors. The BaB process recursively selects an unstable neuron $z_j^{(i)} \notin \mathcal{S}$ to split, generating two child sub-domains to be verified independently. Consequently, the scalability of the verification hinges on the \textbf{branching heuristic}, i.e., the strategy for selecting the optimal neuron to minimize the global relaxation error.

\section{Motivation}

The core efficiency of Branch-and-Bound (BaB) solvers is determined by the refinement strategy used to partition the search space. We observe that existing branching heuristics, despite their empirical success, are fundamentally \textit{witness-agnostic}. In this section, we illustrate why heuristics based on global or average-case information can be misled by specific relaxation bottlenecks and how the CEGAR framework provides a more targeted approach.

\paragraph{The Limitation of Global and Average-case Information.}
Current branching heuristics generally rely on two types of information: (i) static geometric properties, such as the interval width of neurons, or (ii) sensitivity-based estimates that approximate the expected improvement over the entire subdomain. While these metrics provide a useful global view, they fail to account for the local properties of the relaxation at the current failure point. In the context of neural network verification, the lower bound is often blocked by a few "local bottlenecks"—specific input regions where the gap between the linear relaxation and the concrete ReLU function is maximal. Heuristics that aggregate information over the whole domain tend to overlook these critical points, potentially prioritizing neurons that are globally influential but locally tight at the specific counterexample $x^*$ that the solver needs to refute.

\paragraph{A Case Study: The "Wide but Tight" Trap.}
To ground this observation, consider a representative scenario involving two unstable neurons, $n_A$ and $n_B$, with identical output weights ($g_A = g_B = 1$).

Neuron $n_A$ spans a large pre-activation domain $[-2, 18]$ (width 20), while $n_B$ has a significantly narrower domain $[-4, 4]$ (width 8). A heuristic relying on global uncertainty or average sensitivity would naturally prioritize $n_A$ due to its dominant geometric presence.

However, suppose the verifier encounters a spurious counterexample at a witness point $x^*$ where $z_A^* = 8$ and $z_B^* = 0$. At this point, the linear upper bound for $n_A$ is $\hat{z}_A^U = 0.9 z_{pre, A} + 1.8$, which introduces a local relaxation gap of $1.0$. In contrast, the linear upper bound for $n_B$ is $\hat{z}_B^U = 0.5 z_{pre, B} + 2.0$, leading to a gap of $2.0$. Even though $n_B$ is globally "narrower," it is the true source of the spuriousness at $x^*$. Splitting $n_A$ would waste a refinement step on a region that is already locally tight, whereas splitting $n_B$ would eliminate an error twice as large. This example demonstrates that global geometric measures are often decoupled from the local relaxation errors that actually impede the proof.

\paragraph{Branching as Counter-Example Guided Refinement.}
We formalize the solution to this problem by casting the BaB process as a Counter-Example Guided Abstraction Refinement (CEGAR) loop. In this view, any failure to prove safety on the relaxed model yields a \textit{spurious counterexample} $x^*$, which serves as a precise witness to where the abstraction is too coarse. Rather than attempting to reduce the "average" uncertainty of the model, an effective branching strategy should act as a targeted refinement tool. By measuring the \textit{Directional Relaxation Gap} (DRG) specifically at the witness $x^*$, we can prioritize neurons that directly contribute to the current verification failure. This transforms the branching process from a generic space-splitting task into a goal-driven refinement that systematically eliminates spurious witnesses, leading to faster convergence of the safety proof.

\section{Methodology}
\label{sec:methodology}
We propose DRG-BaB, a complete verification framework that reformulates the standard Branch-and-Bound (BaB) search as a Dataflow Counterexample-Guided Abstraction Refinement (CEGAR) loop. Unlike structural CEGAR methods that abstract the network topology, our framework maintains the full network architechture but iteratively refines the abstract domain (the convex relaxation).

\subsection{Framework Overview}
\label{subsec:overview}
We formulate the verification problem as a search for a formal proof of safety within a high-dimensional input domain $\mathcal{D}_0$. To address the NP-completeness of exact reachability analysis, our framework operates a specialized Dataflow CEGAR loop. As outlined in Algorithm~\ref{alg:drg_bab_high_level}, the procedure proceeds through four distinct operational phases for any given sub-domain $\mathcal{D}$ retrieved from the worklist. Crucially, the framework is designed to be fully parallelizable on GPU architectures.
\begin{algorithm}[htbp]
\caption{DRG-BaB: Dataflow CEGAR Verification}
\label{alg:drg_bab_high_level}
\begin{algorithmic}[1]
\Require Neural Network $f$, Input Domain $\mathcal{D}_0$, Property $\mathcal{P}$
\Ensure \textbf{Safe} or \textbf{Unsafe}

\State Initialize worklist $\mathcal{W} \leftarrow \{\mathcal{D}_0\}$

\While{$\mathcal{W} \neq \emptyset$}
    \State Pick and remove a sub-domain $\mathcal{D}$ from $\mathcal{W}$

    \State \texttt{// Step 1: Formal Abstraction}
    \State Compute convex relaxation $\hat{f}$ (linear bounds) over $\mathcal{D}$

    \State \texttt{// Step 2: Safety Check (Verification)}
    \If{Lower bound of $\hat{f}$ satisfies $\mathcal{P}$ over $\mathcal{D}$}
        \State \textbf{continue} \Comment{Abstract proof successful; prune branch}
    \EndIf

    \State \texttt{// Step 3: Counterexample Validation (Falsification)}
    \State Extract candidate counterexample $x^*$ from $\hat{f}$ that violates $\mathcal{P}$
    \If{$f(x^*)$ violates $\mathcal{P}$}
        \Return \textbf{Unsafe} \Comment{Concrete counterexample found}
    \EndIf

    \State \texttt{// Step 4: Dataflow-Driven Refinement}
    \State \textbf{Using spurious witness $x^*$, compute DRG scores}
    \State Select branching neuron $z_k$ to eliminate relaxation error at $x^*$
    \State Split $\mathcal{D}$ into $\{\mathcal{D}_1, \mathcal{D}_2\}$ along $z_k$ and add to $\mathcal{W}$
\EndWhile

\State \Return \textbf{Safe}
\end{algorithmic}
\end{algorithm}

The process begins with \textbf{Formal Abstraction}. Since exact analysis over the non-convex neural surface is computationally intractable, we construct a convex relaxation, denoted as $\hat{f}$, which over-approximates the behavior of the concrete network $f$. In our implementation, we leverage efficient linear bound propagation methods (specifically CROWN~\cite{zhang2018efficient}) to derive symbolic linear lower and upper bounds for the network output. This step effectively transforms the complex non-linear verification problem into a tractable linear computation over the current domain $\mathcal{D}$.

Based on this abstraction, the framework performs a \textbf{Safety Check}. The verifier attempts to certify the safety property $\mathcal{P}$ strictly within the abstract domain by computing the global minimum of the safety margin under the linear relaxation $\hat{f}$. If this abstract lower bound is positive, the soundness of the relaxation guarantees that the concrete network is safe within $\mathcal{D}$. Consequently, the current branch is pruned, successfully terminating the search for this region without further decomposition.

However, if the abstraction is too coarse to prove safety, the underlying solver implicitly identifies a counterexample $x^* \in \mathcal{D}$ that violates $\mathcal{P}$ in the abstract domain. This triggers the \textbf{Counterexample Validation} phase. We treat $x^*$ as a candidate counterexample and immediately evaluate the concrete network function $f(x^*)$. If the concrete output violates the specification, we have identified a genuine error trace; the property is falsified, and the algorithm terminates with an \emph{Unsafe} result.

In cases where $x^*$ violates the property in the abstraction but satisfies it in the concrete system, it is designated as a \emph{spurious counterexample}. This leads to the final phase: \textbf{Dataflow-Driven Refinement}. In the BaB context, this spurious point serves as a precise witness to the local inaccuracy of the convex relaxation. Instead of branching based on static geometric properties, DRG-BaB utilizes $x^*$ to perform a directed sensitivity analysis. By identifying the specific neurons responsible for the relaxation gap at $x^*$, we select an optimal branching variable $z_k$. The domain $\mathcal{D}$ is then split along $z_k$, effectively refining the abstraction to exclude the spurious behavior in the resulting child sub-domains. By repeating this loop, the framework systematically eliminates abstraction errors guided by witnesses of failure, ensuring convergence to either a verified proof or a concrete violation.

\subsection{Identifying the Witness: Spurious Counterexample Generation}\label{subsec:witness_generation}
A prerequisite for our Dataflow-driven refinement is the acquisition of a \emph{spurious counterexample} $x^*$. This point represents the minimizer of the safety margin within the current abstract domain, identifying exactly where the relaxation is loosest relative to the specification. The choice of method to obtain $x^*$ involves a trade-off between geometric precision and computational throughput. We analyze three potential strategies and justify our selection of a closed-form construction.

\paragraph{Option 1: Exact Linear Programming (LP).}
The most rigorous approach is to formulate the search for $x^*$ as a Linear Programming problem over the relaxed polytope. Given the linear constraints defining the convex hull of activation functions, an LP solver can find the exact global minimum of the abstract margin.

However, this approach suffers from a fundamental scalability bottleneck. LP solvers are typically CPU-bound. Integrating them into a GPU-accelerated verifier (like $\alpha$-$\beta$-CROWN) incurs severe overhead due to the incompatible parallel performance between the GPU (where bounds are propagated) and the CPU (where the LP is solved). For high-dimensional networks where thousands of sub-domains are processed in parallel, the latency of exact LP solving renders the branching process computationally intractable.

\paragraph{Option 2: Adversarial Attacks.}
A common alternative is to employ gradient-based adversarial attacks, such as Projected Gradient Descent (PGD), to search for violations. While much faster than LP solving, these methods are fundamentally \textbf{incomplete}. Consequently, if the attack fails to find a counterexample, it provides no mathematical guarantee that the region is safe, making it impossible to prune the branch. More critically for our framework, when an attack fails (i.e., returns a point satisfying the property), it fails to provide a valid "violating" witness $x^*$. Without a counterexample that violates the specification (either concretely or abstractly), the CEGAR loop cannot be meaningfully operated, leaving the branching heuristic without guidance.

\paragraph{Option 3: Closed-Form Construction from Computed Bounds (Ours).}
To overcome the limitations of the above methods, we adopt an analytical approach that leverages the linear bounds already computed during the abstraction phase. Modern efficient verifiers like $\beta$-CROWN utilize dual optimization to propagate a linear lower bound for the margin function, taking the form $m(x) \geq \mathbf{w}^T x + \mathbf{b}$, where $\mathbf{w}$ and $\mathbf{b}$ are derived from the network weights and current neuron bounds.

Since this linear function is a valid lower bound of the relaxation, its minimizer serves as a high-quality proxy for the optimal spurious counterexample. Crucially, minimizing a linear function over a hyper-rectangular input domain $\mathcal{D} = [\mathbf{l}, \mathbf{u}]$ admits a closed-form solution. We construct the witness $x^*$ element-wise by inspecting the sign of the symbolic gradient $\mathbf{w}$:
\begin{equation}
\label{eq:closed_form_witness}
x^*_k =
\begin{cases}
\mathbf{l}_k & \text{if } \mathbf{w}_k \geq 0 \\
\mathbf{u}_k & \text{if } \mathbf{w}_k < 0
\end{cases}
\end{equation}
where $k$ indexes the input dimensions.

\paragraph{Verdict.}
We adopt \textbf{Option 3} for DRG-BaB. This choice offers three decisive advantages. First, unlike adversarial attacks, it is deterministic and aligned with the abstraction: if the bound check fails, Eq.~\ref{eq:closed_form_witness} guarantees the existence of a witness $x^*$ that minimizes the lower bound. Second, it incurs \emph{zero overhead}, as $\mathbf{w}$ is a byproduct of the mandatory bound propagation step. Third, it is fully parallelizable on GPUs, allowing us to generate witnesses for a batch of sub-domains simultaneously without CPU intervention.

%\paragraph{Local Sensitivity Analysis.}
\subsection{Directional Relaxation Gap (DRG) Analysis. }
To effectively refine the abstraction, we must address the \emph{attribution problem}: identifying which unstable neurons are the primary sources of the verification failure. We deconstruct this problem into two components: the \emph{sensitivity} and the \emph{magnitude} of each neuron.

\paragraph{Abstract Sensitivity via Linear Coefficients}\label{secsec:abstract_sensitivity}
A critical design choice in branching heuristics is the definitions of "sensitivity". A straightforward idea would be adopting the gradient of the concrete network, $\nabla_x f(x^*)$. We argue that this approach is suboptimal for CEGAR-based verification.We propose using \textbf{Abstract Sensitivity}, derived directly from the linear coefficients of the relaxation, for three primary reasons.

First, using abstract sensitivity ensures theoretical alignment with the verification objective. Our goal is to refine the \emph{convex relaxation} that erroneously allows the spurious counterexample, rather than to fix the concrete network. Since the spurious witness $x^*$ is the minimizer of the abstract lower bound function $f^L(x) \approx \mathbf{w}^T x + \mathbf{b}$, the propagated linear coefficients $\mathbf{A}_{i,j}$ (representing the dependency of the output on neuron $z_j^{(i)}$) precisely quantify the influence of specific neurons on the current verification bound.

Second, abstract sensitivity prevents information loss in dead zones. For ReLU networks, the concrete gradient $\nabla f$ vanishes if a neuron is inactive at the witness point, creating a blind spot where the heuristic ignores potentially critical neurons simply because they are currently inactive. In contrast, the linear coefficient $\mathbf{A}_{i,j}$ derived from the convex relaxation remains non-zero even in inactive regions, capturing the global potential of the neuron to affect the output via the relaxation envelope.

Third, this approach offers superior computational efficiency. Computing concrete gradients requires additional backward passes through the network graph. Conversely, the linear coefficients $\mathbf{A}$ are a free byproduct of the bound propagation phase in solvers like $\alpha$-$\beta$-CROWN. Utilizing them incurs zero marginal computational cost. Accordingly, we define the abstract sensitivity as $S_{i,j} = |\mathbf{A}_{i,j}|$.

\paragraph{Directional Relaxation Gap. }\label{subsubsec:gap}
Having established sensitivity, we next quantify the relaxation error. While geometrically the relaxation gap exists in two directions (upper and lower bounds), integrating this metric into a dual-optimization verifier reveals a fundamental asymmetry in how these bounds are handled.
\begin{figure}[h]
    \centering
    \begin{subfigure}[b]{0.45\textwidth}
        \centering
        \begin{tikzpicture}[scale=0.75]
            % Axes
            \draw[->] (-2.5,0) -- (2.5,0) node[right] {$z$};
            \draw[->] (0,-1.5) -- (0,2.5) node[above] {$\hat{z}$};
            \draw[thick, black] (-2,0) -- (0,0) -- (2,2);
            \draw[dashed] (-1.5,0) -- (-1.5, 0.2) node[above] {$l$};
            \draw[dashed] (2,0) -- (2, 0.2) node[above] {$u$};
            % Optimized Lower Bound
            \draw[thick, blue, dashed] (-1.5,-0.75) -- (2,1) node[pos=0.1, below, sloped] {\scriptsize Optimized $\alpha$-slope};
            \fill[blue, opacity=0.1] (-1.5,-0.75) -- (0,0) -- (-1.5,0) -- cycle;
            \fill[blue, opacity=0.1] (2.0,2.0) -- (0,0) -- (2.0,1.0) -- cycle;
            \def\zstar{-0.8}
            \def\zrelax{-0.4}
            \node[blue, align=center, font=\scriptsize] at (-1.0, 2.0) {\textbf{Dynamic Lower Bound}\\($y \ge \alpha x$)\\Optimizable Parameter};
            \node[above] at (\zstar, 0) {$z^*$};
            \draw[blue, thick, ->] (\zstar, 0) -- (\zstar, \zrelax) node[midway, left] {$\delta_{L}$};
            %\draw[->, blue, very thick] (1.5, 1.5) -- (1.5, 0.5) node[midway, left] {$\mathbf{A} \ge 0$};
            %\node[blue, align=center, font=\scriptsize] at (1.8, 0.2) {(Pos. Coeff)};
        \end{tikzpicture}
        \caption{Positive Coefficient ($\mathbf{A} \ge 0$):\\ Bound is Adaptive.}
        \label{fig:gap_solver}
    \end{subfigure}
    \hfill
    \begin{subfigure}[b]{0.45\textwidth}
        \centering
        \begin{tikzpicture}[scale=0.75]
            % Axes
            \draw[->] (-2.5,0) -- (2.5,0) node[right] {$z$};
            \draw[->] (0,-1.5) -- (0,2.5) node[above] {$\hat{z}$};
            \draw[thick, black] (-2,0) -- (0,0) -- (2,2);
            \draw[dashed] (-1.5,0) -- (-1.5, -0.2) node[below] {$l$};
            \draw[dashed] (2,0) -- (2, -0.2) node[below] {$u$};
            % Static Upper Bound
            \draw[thick, red, dashed] (-1.5,0) -- (2,2) node[pos=0.2, above, sloped] {\scriptsize Static Triangle};
            \fill[red, opacity=0.1] (-1.5,0) -- (2,2) -- (0,0) -- cycle;

            \def\zstar{-0.8}
            \def\zrelax{0.45}
            %\draw[dotted, thick] (\zstar, 0) -- (\zstar, \zrelax);
            \node[below] at (\zstar, 0) {$z^*$};
            \draw[red, thick, ->] (\zstar, 0) -- (\zstar, \zrelax) node[midway, right] {$\delta_{U}$};

            \node[red, align=center, font=\scriptsize] at (1.2, -0.8) {\textbf{Static Upper Bound}\\Requires Branching};

            % Gradient Arrow (Positive)
            %\draw[->, red, very thick] (1.5, 0.5) -- (1.5, 1.5) node[midway, right] {$\mathbf{A} < 0$};
            %\node[red, align=center, font=\scriptsize] at (2.0, 0.2) {(Neg. Coeff)};
        \end{tikzpicture}
        \caption{Negative Coefficient ($\mathbf{A} < 0$):\\ Bound is Static.}
        \label{fig:gap_branching}
    \end{subfigure}
    \caption{\textbf{Visualizing the Intuition of Solver-Aware Branching.} (a) When $\mathbf{A} \ge 0$, the lower bound is active. In $\alpha$-CROWN, this bound is defined by an adjustable slope $\alpha$, allowing the solver to dynamically reduce the error. (b) When $\mathbf{A} < 0$, the upper bound is active. This bound is the McCormick triangle, which is structurally static. Branching is the only mechanism to tighten this rigid geometry.}
    \label{fig:orthogonal_gap}
\end{figure}
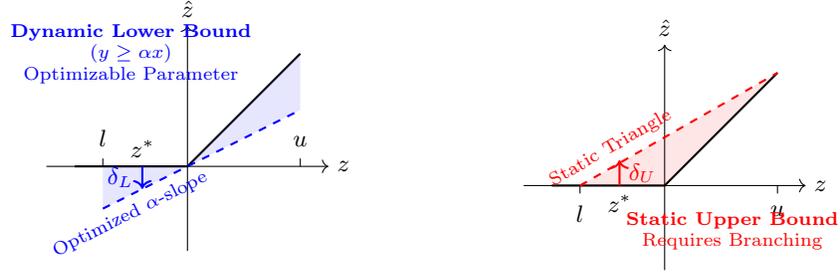

As visualized in Fig.~\ref{fig:orthogonal_gap}, the verifier handles the two relaxation bounds differently based on the sign of the linear coefficient. When the linear coefficient $\mathbf{A}_{i,j}$ is positive, the solver minimizes the margin by exploiting the lower bound of the ReLU. In $\alpha$-CROWN, this lower bound is parameterized as $y \ge \alpha x$, where $\alpha$ is a variable optimized via dual ascent. Since the solver can continuously adjust $\alpha$ to tighten the bound, the marginal benefit of discrete branching in this direction is significantly reduced, as the error is adaptive.

Conversely, when $\mathbf{A}_{i,j}$ is negative, the solver exploits the linear upper bound. This bound corresponds to the standard McCormick triangle connecting $(l,0)$ and $(u,u)$. Unlike the lower bound, the geometry of this triangle is static and determined solely by the domain endpoints. The dual optimizer cannot alter this shape.

Based on this observation, we propose a \textbf{Solver-Aware Refinement Strategy}. The intuition is to allocate the branching budget to the structural bottlenecks where the solver is powerless. Since the upper bound is rigid and impervious to optimization, we hypothesize that branching on neurons dominated by upper-bound errors yields a higher return on investment. Therefore, we restrict our gap metric to the static upper-bound violation, defined as:
\begin{equation}
\label{eq:directional_gap_asym}
\delta_{orth}(z^*) = \mathbb{I}(\mathbf{A}_{i,j} < 0) \cdot \left( \hat{z}^U(z^*) - \text{ReLU}(z^*) \right)
\end{equation}
where $\hat{z}^U$ is the linear upper bound value at the witness.

\subsubsection{The DRG Branching Score}
\label{subsubsec:score}

Finally, we synthesize the abstract sensitivity and the solver-aware gap into a unified branching score. The score for neuron $j$ in layer $i$ is defined as the product of its contribution to the abstract bound and its uncorrectable relaxation error:
\begin{equation}
\label{eq:final_score}
Score_{i,j} = \underbrace{|\mathbf{A}_{i,j}|}_{\text{Sensitivity}} \times \underbrace{\delta_{orth}(z_j^{(i)}(x^*))}_{\text{Static Error}}
\end{equation}
By prioritizing neurons with high scores, DRG-BaB effectively targets the specific regions of the abstraction that are both influential to the safety property and geometrically loose in a way that the continuous optimizer cannot resolve.

\section{Experiments}
\label{sec:experiments}

In this section, we present a comprehensive evaluation of our proposed framework, \textbf{DRG-BaB}, aiming to validate its efficiency, effectiveness, and scalability in verifying neural network robustness. By integrating our heuristic into the state-of-the-art verifier \textbf{$\alpha$-$\beta$-CROWN}, our experimental design is structured to answer the following three key Research Questions (RQs):

\noindent\textbf{RQ1: Comparative Scalability on Hard Instances.}
Does DRG-BaB improve the success rate and runtime performance on challenging high-dimensional benchmarks (e.g., CIFAR-10) compared to established SOTA heuristics like BaBSR? This question assesses the overall practical value of the framework.

\noindent\textbf{RQ2: Component Decomposition and Ablation.}
How do the distinct components of our metric, specifically the use of \textit{abstract sensitivity} and the \textit{spurious counterexample}, contribute to the overall guidance efficacy? We disentangle these factors to verify that identifying the precise location of the failure is a prerequisite for effective branching.

\noindent\textbf{RQ3: Verification of the Orthogonality Hypothesis.}
Does the experimental evidence support our theoretical insight regarding the asymmetry of optimization? Specifically, we investigate whether masking the dynamic lower-bound errors (which are handled by the solver) and focusing solely on the static upper-bound errors (via our single-sided metric) yields a superior Return on Investment (ROI) compared to a symmetric approach.

\subsection{Experimental Setup}
\label{subsec:setup}

\noindent\textbf{Benchmarks and Rationale.}
We evaluate our framework on two standard image classification benchmarks: \textbf{MNIST} and \textbf{CIFAR-10}. We deliberately focus on these high-dimensional verification tasks rather than low-dimensional control benchmarks (e.g., ACAS Xu). The rationale lies in the fundamental nature of the search space: while \textit{input domain splitting} is effective for low-dimensional problems, it becomes computationally intractable for image inputs due to the curse of dimensionality. Consequently, scaling verification to the image domain necessitates \textit{intermediate neuron splitting} (activation branching). By selecting these benchmarks, we rigorously evaluate the ability of our heuristic to navigate the complex, non-convex activation space—a challenge that cannot be bypassed by simple input partitioning. Specific network architectures and perturbation radii ($\epsilon$) are detailed in the respective Research Questions.

\noindent\textbf{Baselines.}
We categorize branching heuristics into two distinct classes based on their computational overhead:
\begin{itemize}
    \item \textbf{One-Shot Heuristics:} These methods compute branching scores using only currently available information (e.g., bounds or gradients) without performing trial-and-error simulations.
    \begin{itemize}
        % \item \textbf{Random:} Selects neurons uniformly at random.
        % \item \textbf{Intercept:} Selects neurons with the largest intercetp.
        \item \textbf{BaBSR~\cite{bunel2020branch}:} The standard gradient-based heuristic. It serves as our \textbf{primary baseline} as it shares the same zero-overhead characteristic as our method.
    \end{itemize}

    \item \textbf{Lookahead Heuristics:} These methods simulate branches to estimate exact gains.
    \begin{itemize}
        \item \textbf{FSB~\cite{wang2021beta}:} The state-of-the-art lookahead heuristic. Note that FSB incurs significantly higher computational cost and memory usage (due to batch expansion) compared to one-shot methods. We include FSB not as a direct competitor for efficiency, but as a \textbf{high-water mark reference} to quantify how close our analytical metric can approach the quality of exhaustive simulation.
    \end{itemize}
\end{itemize}

\noindent\textbf{Implementation and Environment.}
We integrated the DRG heuristic directly into \textbf{$\alpha$-$\beta$-CROWN}~\cite{wang2021beta}, the state-of-the-art complete verifier. Crucially, strictly adhering to the methodology in Sec.~\ref{subsec:witness_generation}, our implementation avoids any calls to external linear solvers (e.g., Gurobi or Scipy). The computation of spurious witnesses and DRG scores is fully vectorized using PyTorch, enabling massive parallelism across thousands of sub-domains on the GPU. All experiments were conducted on a server equipped with an Intel Xeon E5-2678 v3 CPU and a single NVIDIA RTX 4090 GPU.

\subsection{RQ1: Overall Performance}

\begin{table}[t]
\centering
\caption{\textbf{Overall Performance.}
We report the mean and median (in parentheses) for branches and runtime.
The \textbf{Win Rate} reflects the percentage of instances where DRG is better or equal ($\le$) to BaBSR.
\textbf{FSB} is included as a lookahead-based reference. }
\label{tab:rq1_results}
\setlength{\tabcolsep}{3pt}
\renewcommand{\arraystretch}{1.1}
\resizebox{\textwidth}{!}{%
\begin{tabular}{l|l|rr|rr|r|cc}
\toprule
\multirow{2}{*}{\textbf{Benchmark}} & \multirow{2}{*}{\textbf{Method}} & \multicolumn{2}{c|}{\textbf{Branches}} & \multicolumn{2}{c|}{\textbf{Time (s)}} & \multirow{2}{*}{\textbf{\%Timeout}} & \multicolumn{2}{c}{\textbf{Win Rate (\%)}} \\
\cmidrule(lr){3-4} \cmidrule(lr){5-6} \cmidrule(lr){8-9}
& & \textbf{Mean} & \textbf{(Median)} & \textbf{Mean} & \textbf{(Median)} & & \textbf{Br. ($\le$)} & \textbf{Time ($\le$)} \\
\midrule

% --- oval_deep ---
\multirow{3}{*}{\shortstack[l]{CIFAR-10\\Deep}}
 & BaBSR & 1,699 & (1) & 9.80 & (8.12) & 0\% & -- & -- \\
 & \textbf{DRG (Ours)} & \textbf{8} & \textbf{(0)} & \textbf{7.07} & \textbf{(7.04)} & 0\% & \textbf{93} & \textbf{98} \\
 \cmidrule{2-9}
 & \textit{FSB (Ref.)} & \textit{6} & \textit{(0)} & \textit{10.27} & \textit{(10.29)} & \textit{0\%} & -- & -- \\
 \midrule

% --- oval_wide ---
\multirow{3}{*}{\shortstack[l]{CIFAR-10\\Wide}}
 & BaBSR & \textbf{33,533} & (5) & \textbf{25.18} & (4.39) & 3\% & -- & -- \\
 & \textbf{DRG (Ours)} & 52,884 & \textbf{(4)} & 30.90 & \textbf{(3.04)} & 4\% & \textbf{83} & \textbf{93} \\
 \cmidrule{2-9}
 & \textit{FSB (Ref.)} & \textit{19,768} & \textit{(3)} & \textit{19.61} & \textit{(3.43)} & \textit{2\%} & -- & -- \\
 \midrule

% --- oval_base ---
\multirow{3}{*}{\shortstack[l]{CIFAR-10\\Base}}
 & BaBSR & 63,474 & (52) & 40.66 & (3.93) & 5\% & -- & -- \\
 & \textbf{DRG (Ours)} & \textbf{48,757} & \textbf{(50)} & \textbf{29.37} & \textbf{(3.61)} & 4\% & \textbf{60} & \textbf{82} \\
 \cmidrule{2-9}
 & \textit{FSB (Ref.)} & \textit{34,936} & \textit{(32)} & \textit{26.11} & \textit{(4.37)} & \textit{3\%} & -- & -- \\
 \midrule

% --- mnist_conv_big ---
\multirow{3}{*}{\shortstack[l]{MNIST\\ConvBig}}
 & BaBSR & \textbf{911} & \textbf{(0)} & \textbf{8.53} & \textbf{(0.62)} & 1\% & -- & -- \\
 & \textbf{DRG (Ours)} & 996 & \textbf{(0)} & 9.64 & (0.62) & 1\% & \textbf{99} & \textbf{44} \\
 \cmidrule{2-9}
 & \textit{FSB (Ref.)} & \textit{651} & \textit{(0)} & \textit{8.40} & \textit{(0.72)} & \textit{1\%} & -- & -- \\

\bottomrule
\end{tabular}%
}
\end{table}
We evaluate the performance of DRG-BaB against the standard improvement-estimation heuristic, BaBSR, with the lookahead-based FSB serving as a performance reference. Table~\ref{tab:rq1_results} summarizes the results across four benchmarks, focusing on how different branching strategies navigate the search space across varying network architectures.

On the \textit{CIFAR-10 Deep} benchmark, we observe a significant divergence in branching efficiency. While BaBSR requires an average of 1,699 branches, DRG-BaB reduces this to a mean of 8, achieving a 98\% win rate in runtime. The median branch count of 0 for DRG-BaB indicates that the majority of these deep-network properties are resolved at or very near the root node. This suggests that the witness-guided strategy is particularly effective for deeper architectures, where identifying the specific relaxation error responsible for the proof failure provides more decisive guidance than estimating potential bound improvements. Notably, on this specific model, the analytical guidance of DRG-BaB results in a lower mean runtime than even the simulation-based FSB (7.07s vs. 10.27s).

The results for the \textit{CIFAR-10 Wide} and \textit{Base} models exhibit a more complex distribution. On the \textit{Wide} model, DRG-BaB achieves a superior median branch count (4 vs. 5) and a high win rate of over 80\% in both branches and time. However, the mean branch count for DRG-BaB is higher than that of BaBSR. This discrepancy indicates a skewed distribution where DRG-BaB excels in the vast majority of cases but encounters a small number of difficult outliers that require significantly larger search trees. On the \textit{Base} model, the performance is more consistent, with DRG-BaB outperforming BaBSR in mean branches, median branches, and overall runtime. These results demonstrate that while improvement estimation is competitive on shallower or wider layers, the goal-driven refinement of DRG-BaB generally leads to more efficient conflict resolution.

On the \textit{MNIST ConvBig} benchmark, the verification task is relatively trivial for all heuristics, as evidenced by the median branch count of 0 across all methods. In fact, we notice that most of the cases within can be easily dealed with pgd or incomplete verifier.
In this regime, the branching quality is nearly identical, with a branch-wise win rate of 99\% (primarily consisting of ties at the root node). However, the runtime win rate for DRG-BaB drops to 44\%. This slight increase in mean runtime reflects the additional computational steps required to analytically construct the spurious witness and calculate the relaxation gap. For models that are easily solved at the root node, this marginal complexity does not translate into further branch reduction, resulting in a slight runtime penalty compared to the simpler logic of BaBSR.

In comparison with the lookahead-based reference (FSB), DRG-BaB represents a middle ground in the performance landscape. While FSB typically achieves the lowest branch counts by simulating potential splits, it often requires more total time on simpler or deeper models due to its higher per-node complexity. DRG-BaB achieves branching quality that is significantly better than estimation-based methods and approaches the performance of simulation-based methods, particularly in instances where the spurious counterexample provides a clear signal for refinement. This confirms that the directional relaxation gap is a robust metric for guiding the branching process across diverse neural architectures.

\subsection{RQ2: Component Decomposition and Ablation}
\label{subsec:rq2_ablation}
To validate the design choices of our branching metric, we perform a controlled ablation study on the components of the proposed score: $Score_{i,j} = |\mathbf{A}_{i,j}| \times \delta_{orth}(z_j^{(i)}(x^*))$. We isolate the contribution of \textit{Abstract Sensitivity} ($|\mathbf{A}|$) and \textit{Witness Location} ($x^*$) by evaluating three variants: \textbf{Variant-Grad} (uses concrete gradients via backpropagation), \textbf{Variant-Center} (calculates gaps at the geometric domain center), and \textbf{Variant-Intercept} (uses the relaxation intercept as a location-agnostic guidance signal).
% To validate the design choices of our branching metric, we perform a controlled ablation study. Our proposed score is defined as $Score_{i,j} = |\mathbf{A}_{i,j}| \times \delta_{orth}(z_j^{(i)}(x^*))$. We isolate the contribution of the \textit{Abstract Sensitivity} ($|\mathbf{A}|$) and the \textit{Witness Location} ($x^*$) by evaluating the following variants against our full method:

% \begin{itemize}
%     \item \textbf{Variant-Grad (Concrete Sensitivity):}
%     Replaces the linear coefficient $|\mathbf{A}_{i,j}|$ with the gradient of the concrete network loss $|\nabla_{z} \mathcal{L}(x^*)|$ computed via backpropagation.
%     \item \textbf{Variant-Center (Location Agnostic):}
%     Replaces the spurious witness $x^*$ with the geometric center of the current domain.
%     \item \textbf{Variant-Intercept (Location Agnostic):}
%     Replaces the gap computed using spurious witness $x^*$ with the intercept.
% \end{itemize}

\begin{table}[h]
\centering
\caption{\textbf{RQ2: Model Component Ablation.}
We evaluate the impact of sensitivity source and guidance location on both Deep and Wide architectures.
Branches are reported as Mean (Median).}
\label{tab:rq2_ablation_multi}
\resizebox{\textwidth}{!}{%
\begin{tabular}{l|cc|cc|cc}
\toprule
\multirow{2}{*}{\textbf{Method / Variant}} & \multicolumn{2}{c|}{\textbf{Sensitivity \& Guidance}} & \multicolumn{2}{c|}{\textbf{CIFAR-10 Deep}} & \multicolumn{2}{c}{\textbf{CIFAR-10 Wide}} \\
& \textbf{Sens.} & \textbf{Loc.} & \textbf{Branches} & \textbf{Time (s)} & \textbf{Branches} & \textbf{Time (s)} \\
\midrule
Variant-Grad & Concrete $\nabla \mathcal{L}$ & Witness $x^*$& 310939 (44) & 212.1  & 695209 (984787) & 398.0\\
Variant-Center & Abstract $\mathbf{A}$ & Center $x_c$ & 11 (0) & 7.5 & 79847 (4) & 45.1 \\
Variant-Intercept & Abstract $\mathbf{A}$ & Intercept & 11385 (0) & 14.9 & 96097 (3) & 55.2 \\
\midrule
\textbf{Ours (DRG)} & \textbf{Abstract $\mathbf{A}$} & \textbf{Witness $x^*$} & \textbf{8 (0)} & \textbf{7.1} & \textbf{52884 (4)} & \textbf{30.9} \\
\bottomrule
\end{tabular}%
}
\end{table}
Table~\ref{tab:rq2_ablation_multi} decomposes the performance impact of our two core design components: Abstract Sensitivity ($\mathbf{A}$) and Witness Guidance ($x^*$). The results reveal distinct failure modes when either component is replaced.

The impact of sensitivity source is most evident when comparing Variant-Grad with our full method. Variant-Grad replaces abstract sensitivity with concrete gradients computed at the witness point. As shown in the table, this variant leads to a substantial increase in both branch counts and verification time. For the CIFAR-10 Deep architecture, the branch count rises from 8 to over 300,000. This confirms that in deep architectures, backpropagated signals are prone to vanishing or noise. In contrast, the linear coefficients $\mathbf{A}$ derived from the global relaxation provide robust guidance that penetrates network depth.

The results for Variant-Center and Variant-Intercept demonstrate the necessity of using the spurious witness $x^*$ to guide the branching process. While Variant-Center performs reasonably well on the Deep architecture, its performance degrades on the Wide architecture, requiring approximately 50\% more branches than our method (79,847 vs. 52,884). This indicates that splitting at the geometric center is a useful baseline but fails to account for the specific areas of the input domain that cause property violations. Similarly, using the intercept as a location-agnostic signal yields sub-optimal results on larger models. The intercept provides a coarse estimate of the relaxation's bias but lacks the precise spatial information provided by the witness $x^*$.

The data suggests a clear synergy between abstract sensitivity and witness-based guidance. The failure of Variant-Grad confirms that the branching heuristic must be tightly coupled with the underlying abstraction (e.g., CROWN or $\alpha,\beta$-CROWN) rather than the original concrete network. Furthermore, the performance gap between Variant-Center and our method indicates that the "trouble spots" identified by the solver (the witnesses) are more effective targets for domain refinement than the center of the domain.

\subsection{RQ3: Validation of the Orthogonality Hypothesis}
\label{subsec:rq3_mechanism}
In Sec.~\ref{subsubsec:gap}, we hypothesized that due to the asymmetry of $\alpha$-CROWN's dual optimization, branching should focus specifically on the static upper-bound errors. To verify this \textit{Orthogonality Hypothesis}, we compare our single-sided metric against a symmetric baseline:

\begin{itemize}
    \item \textbf{Variant-Sym (Symmetric Gap):}
    Calculates the relaxation gap in both directions, regardless of the solver's optimization capability. The score is defined as follows:
    \begin{equation*}
        |\mathbf{A}_{i,j}| \times (\mathbb{I}(\mathbf{A}_{i,j} < 0) \cdot \left( \hat{z}^U(z^*) - \text{ReLU}(z^*) \right) + \mathbb{I}(\mathbf{A}_{i,j} \ge 0) \cdot \left( \text{ReLU}(z^*) -\hat{z}^L(z^*)   \right) )
    \end{equation*}

    \item \textbf{Ours (Asymmetric/Orthogonal Gap):}
    Masks the lower-bound error, focusing strictly on the static upper-bound error where the solver is powerless (See Eq.~\ref{eq:final_score}).
\end{itemize}

\begin{figure*}[h]
    \centering
    % 第一张图：CIFAR-10 Base
    \begin{subfigure}[b]{0.32\textwidth}
        \centering
        \includegraphics[width=\linewidth]{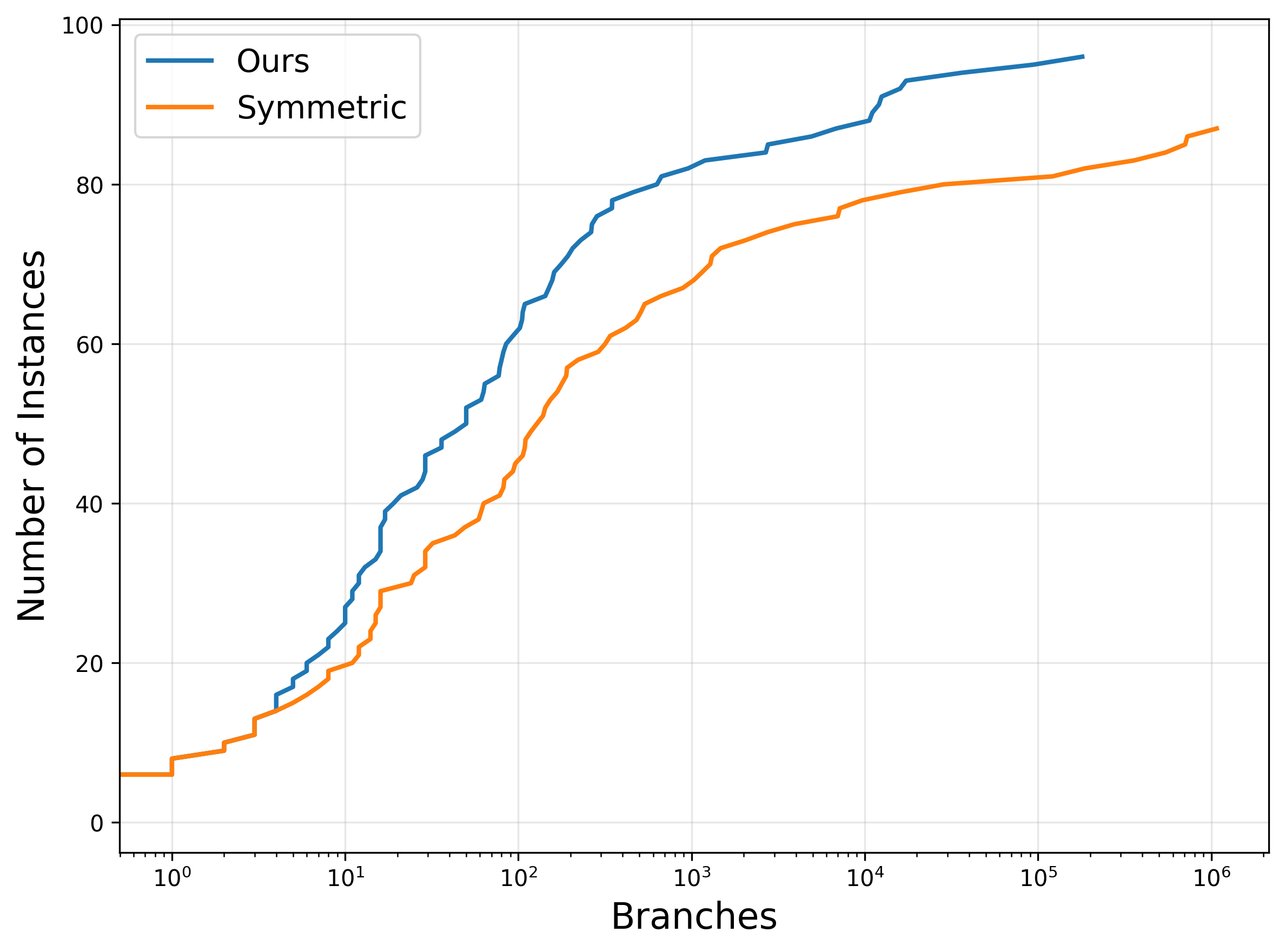} % 替换为你的文件名
        \caption{CIFAR-10 Base}
        \label{fig:cactus_base}
    \end{subfigure}
    \hfill
    % 第二张图：CIFAR-10 Wide
    \begin{subfigure}[b]{0.32\textwidth}
        \centering
        \includegraphics[width=\linewidth]{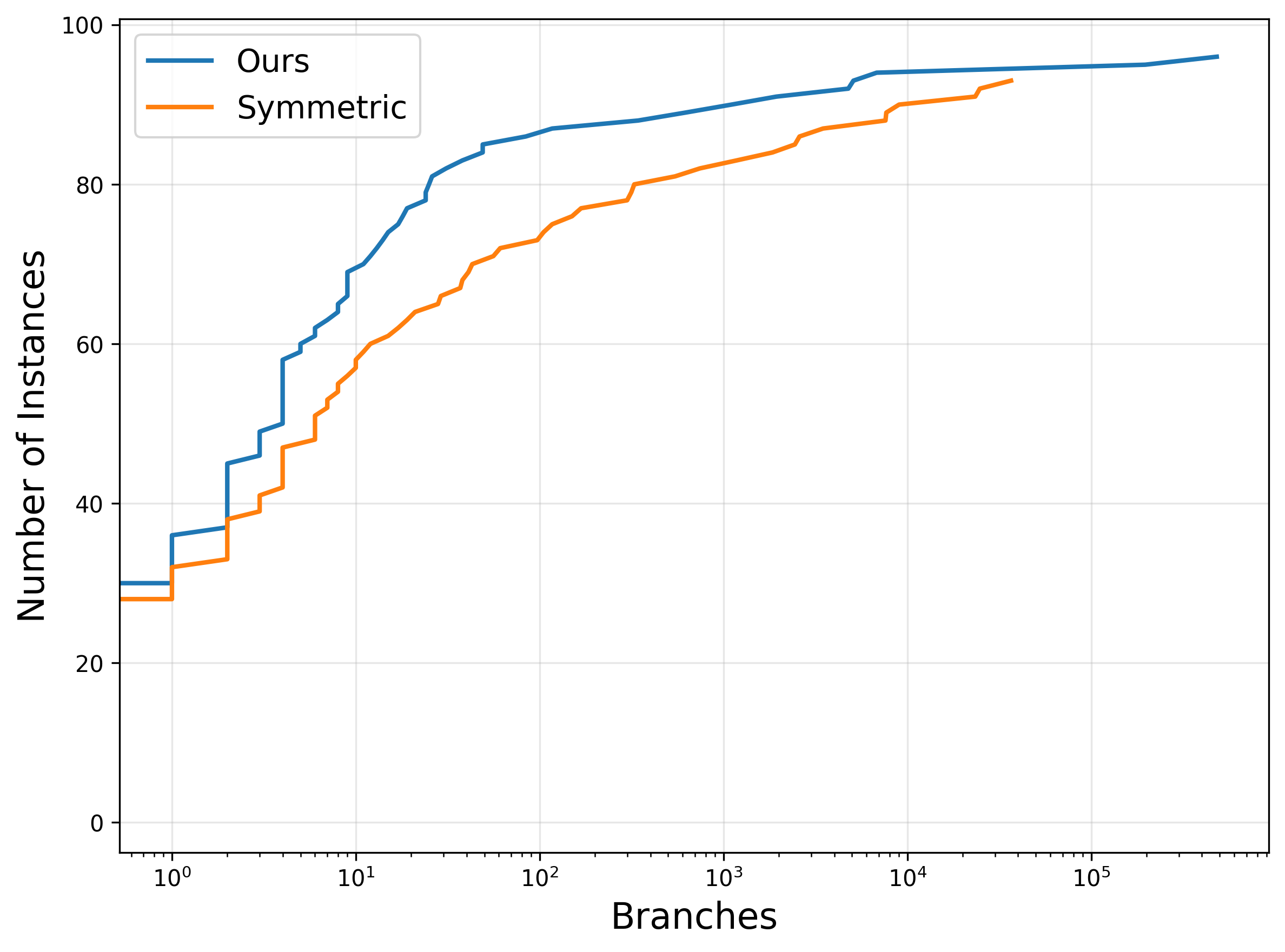} % 替换为你的文件名
        \caption{CIFAR-10 Wide}
        \label{fig:cactus_wide}
    \end{subfigure}
    \hfill
    % 第三张图：CIFAR-10 Deep
    \begin{subfigure}[b]{0.32\textwidth}
        \centering
        \includegraphics[width=\linewidth]{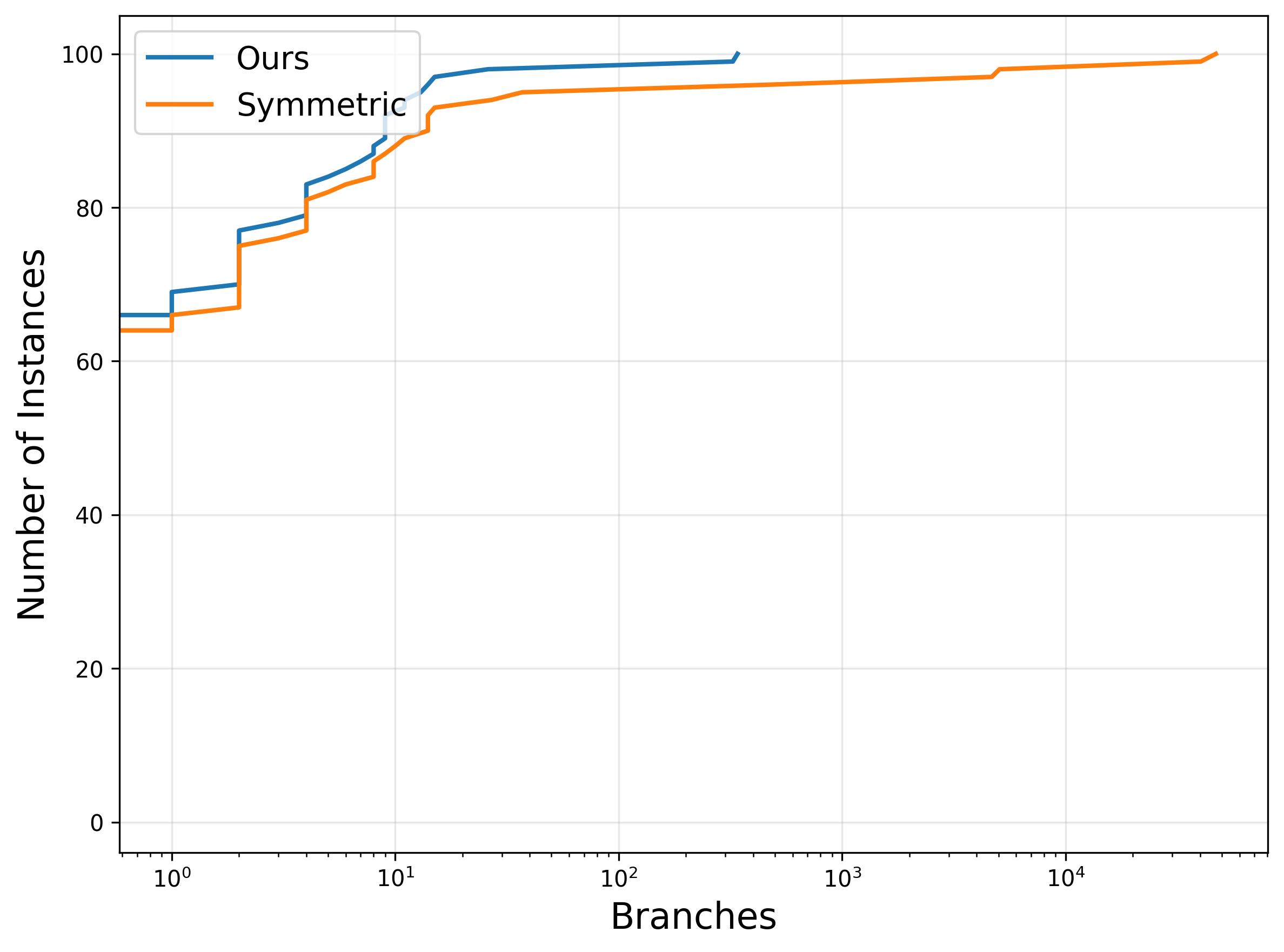} % 替换为你的文件名
        \caption{CIFAR-10 Deep}
        \label{fig:cactus_deep}
    \end{subfigure}

    \caption{\textbf{Verification of the Orthogonality Hypothesis (RQ3).}
    The cactus plots show the cumulative number of solved instances versus the number of visited branches (log scale) for different CIFAR-10 architectures.}
    \label{fig:cactus_rq3_main}
\end{figure*}

\begin{table}[h]
    \centering
    \caption{\textbf{Quantitative Verification of the Orthogonality Hypothesis (RQ3).} We compare the branching efficiency of the Symmetric metric versus our Asymmetric metric (Upper Gap only). \textbf{Avg. Branches} is calculated over instances solved by both methods. \textbf{Time} indicates total verification time. The significant reduction in branches confirming that masking the dynamic lower-bound errors aligns better with the solver's optimization capability.}
    \label{tab:rq3_aggregate}
    \setlength{\tabcolsep}{6pt}
    \resizebox{\textwidth}{!}{
    \begin{tabular}{l|cc|cc|cc}
        \toprule
        \multirow{2}{*}{\textbf{Benchmark}} & \multicolumn{2}{c|}{\textbf{Symmetric}} & \multicolumn{2}{c|}{\textbf{Ours}} & \multicolumn{2}{c}{\textbf{Improvement}} \\
        & Branches & Time (s) & Branches & Time (s) & \textbf{Branch Red.} & \textbf{Time Red.} \\
        \midrule
        CIFAR-10 Base & 43855 & 9965 & \textbf{578} & \textbf{2939} & \textbf{98.7\%} & \textbf{70.5\%} \\
        CIFAR-10 Wide & 1334 & 4614 & \textbf{227} & \textbf{3092} & \textbf{83.0\%} & \textbf{33.0\%} \\
        CIFAR-10 Deep & 976 & 790 & \textbf{9} & \textbf{707} & \textbf{99.1\%} & \textbf{10.5\%} \\
        \midrule
        \textbf{Overall} & \textbf{1149} & \textbf{5405} & \textbf{114} & \textbf{3799} & \textbf{90.1\%} & \textbf{29.7\%} \\
        \bottomrule
    \end{tabular}
    }
\end{table}

\begin{table}[h]
    \centering
    \caption{\textbf{Head-to-Head Comparison: Symmetric vs. Asymmetric.} We categorize instances based on whether our approach requires fewer (Win), equal (Tie), or more (Loss) branches than the Symmetric baseline. Only instances where both methods are unknown are excluded.}
    \label{tab:rq3_head_to_head}
    \setlength{\tabcolsep}{10pt}
    \begin{tabular}{l c c c c}
        \toprule
        \textbf{Benchmark} & \textbf{Ours. Wins} & \textbf{Ties} & \textbf{Sym. Wins} & \textbf{Dom. Rate} \\
        & (Fewer Br.) & (Equal) & (Fewer Br.) &  \\
        \midrule
        CIFAR-10 Base & 61 & 34 & 1 & \textbf{98.4\%} \\
        CIFAR-10 Wide & 49 & 47 & 0 & \textbf{100\%} \\
        CIFAR-10 Deep & 14 & 86 & 0 & \textbf{100\%} \\
        \midrule
        \textbf{Total} & \textbf{63} & \textbf{133} & \textbf{0} & \textbf{100\%} \\
        \bottomrule
    \end{tabular}
\end{table}
We evaluate the validity of the Orthogonality Hypothesis by comparing the performance of our proposed asymmetric metric against the symmetric baseline across CIFAR-10 benchmarks. The visual performance profiles are presented in the cactus plots (Fig.~\ref{fig:cactus_rq3_main}), and the quantitative data is detailed in Table~\ref{tab:rq3_aggregate} and Table~\ref{tab:rq3_head_to_head}.

The cactus plots reveal a clear performance separation between the two methods. On all three benchmarks, the curve representing our asymmetric metric (Ours) consistently remains above and to the left of the symmetric baseline, indicating that it solves more instances within any given budget of branches or time. This gap is quantitatively confirmed in Table~ref{tab:rq3\_aggregate}. The most striking result is observed on the \textit{CIFAR-10 Base} benchmark, where masking the lower-bound error reduces the average number of branches by over 98\%, dropping from 43,855 to 578. Similar trends are evident on \textit{Wide} and \textit{Deep} architectures, with branch reductions of 83.0\% and 99.1\%, respectively. These orders-of-magnitude reductions suggest that the lower-bound relaxation error, which is actively minimized by the solver's dual optimization, acts as heuristic noise. Including this redundant information in the branching score dilutes the signal from the static upper-bound errors, misleading the search into unnecessary sub-domains. By filtering out this noise, our asymmetric metric directs the verification process toward the actual structural bottlenecks.

Table~\ref{tab:rq3_head_to_head} provides a granular head-to-head comparison, categorizing instances based on which method requires fewer branches. The results demonstrate that the asymmetric metric is strictly superior or equal to the symmetric baseline in almost every case. Across all tested benchmarks, the symmetric metric wins in only a single instance (on \textit{Base}), resulting in a dominance rate of effectively 100\%. A significant portion of the results are ties (e.g., 86 ties on \textit{Deep}), which typically correspond to trivial instances solved at the root node or within very few branches where both heuristics perform identically. However, in non-trivial cases where branching is required, our method consistently finds a more efficient proof. This confirms that removing the lower-bound component does not lead to information loss; rather, it removes a confounding factor that systematically degrades decision quality.

The experimental evidence strongly supports our hypothesis regarding the interplay between the solver and the branching heuristic. In the $\alpha$-$\beta$-CROWN framework, the lower bound is dynamic and adaptive, continuously tightened by optimizing the parameters. Conversely, the upper bound (the McCormick triangle) is static and geometric. The results show that branching resources yield a significantly higher return on investment when allocated exclusively to the static errors that the optimizer cannot resolve. The fact that performance improves drastically when half of the "error information" (the lower gap) is ignored indicates that the optimization and branching processes should be treated as orthogonal mechanisms. The optimizer handles the lower bound, while the branching heuristic handles the upper bound. Respecting this division of labor is crucial for maximizing the efficiency of complete neural network verification.

\section{Related Work}
The verification of DNNs has evolved from general-purpose solvers to specialized frameworks that balance theoretical soundness with computational efficiency.
\paragraph{Complete Verification Paradigms.} Early complete verifiers formulated the problem as a constraint satisfiability task. Reluplex~\cite{katz2017reluplex} adapted the Simplex algorithm to handle ReLU constraints, marking the first scalable SMT-based approach and laying the foundation for the \textbf{Marabou} framework~\cite{katz2019marabou,wu2024marabou}. While Mixed-Integer Programming (MIP) solvers~\cite{tjeng2017verifying,kouvaros2018formal} provide rigorous bounds, they often struggle with the NP-complete state space inherent to large-scale networks.
Modern verifiers~\cite{bak2021nnenum,henriksen2020efficient,wang2018neurify} combine symbolic interval propagation with iterative domain splitting. Recent breakthroughs have moved toward GPU-accelerated bound propagation. This is exemplified by BaDNB~\cite{depalma2021improved} and \textbf{$\alpha$-$\beta$-CROWN}~\cite{wang2021beta}, which exploit massive parallelism to resolve thousands of sub-problems concurrently. Our work operates within this high-throughput BaB paradigm, specifically aiming to optimize the critical branching decisions that drive convergence.

\paragraph{Linear Relaxation and Abstraction.} The efficiency of any BaB-based approach depends heavily on the precision of its underlying abstract domain. Linear relaxation techniques~\cite{singh2019deep,singh2018fast,gehr2018ai2,wong2018provable,zhang2018efficient} over-approximate non-linear activations with polyhedral constraints to compute sound lower bounds. These can be interpreted as \textit{dataflow abstractions} of the concrete network. While multi-neuron constraints~\cite{muller2022prima,ferrari2022complete} offer tighter abstractions by capturing inter-neuron dependencies, they introduce substantial computational overhead. In contrast, our approach is rooted in $\alpha$-$\beta$-CROWN~\cite{wang2021beta}, preserving the efficiency of simpler relaxations while dynamically recovering precision through targeted branching in regions where abstraction error most severely hinders the proof.

\paragraph{Branching Heuristics.} To address scalability, the community has largely converged on the \textit{Branch-and-Bound} (BaB) framework~\cite{bunel2018unified,depalma2021improved,henriksen2021deepsplit,kouvaros2021towards,lu2020neural,shi2022efficiently,wang2021beta,xu2021fast}, where branching serves as the primary \textit{refinement} mechanism. \textit{Geometric strategies}, such as layer-wise splitting~\cite{yin2022layersar} or input partitioning~\cite{wang2018formal}, prioritize intervals based on width. However, interval width is often a poor proxy for the actual difficulty of a verification sub-problem. Alternatively, \textit{improvement-estimation methods} attempt to predict the gain in bound tightening. BaBSR~\cite{bunel2020branch} utilizes gradient-based sensitivity, while FSB~\cite{wang2021beta} employs a more intensive lookahead procedure. Learning-based heuristics~\cite{lu2020neural}, train Graph Neural Networks to predict optimal splits. Despite their sophistication, these methods generally rely on average-case sensitivity or static properties. They fail to explicitly leverage the \textit{witness of failure} (the spurious counterexample) generated during the verification process, which we argue is the most direct and effective guide for refinement. Recently, \textbf{GenBaB}~\cite{shi2025neural} extended this paradigm beyond ReLU networks, proposing a general framework for verifying diverse nonlinearities (e.g., Sigmoid, Transformer attention) via pre-optimized bounding tables. While GenBaB expands the \textit{scope} of BaB to general computation graphs, our work focuses on optimizing the \textit{branching decision} itself. We aim to accelerate convergence by exploiting the geometric information inherent in verification failures, a strategy compatible with both ReLU and general nonlinear architectures.

\paragraph{CEGAR in DNN Verification.}
Our framework is conceptually grounded in \textit{CounterExample Guided Abstraction Refinement} (CEGAR)~\cite{clarke2000counterexample}. In the DNN context, CEGAR has traditionally been applied to \textbf{structural abstraction}~\cite{elboher2020abstraction,liu2024abstraction,demarchi2022counterexample,zhao2022cleverest,antal2025counterexample}, which construct a smaller, over-approximated version of the original network to accelerate verification. In these approaches, once a spurious counterexample $x^*$ is identified, it is used to refine the abstract model—essentially creating a more precise over-approximation that is no longer falsified by $x^*$. However, the potential of using CEGAR to guide internal neuron branching remains significantly under-explored. Unlike existing heuristics that branch "blindly" based on average-case sensitivity, our approach treats the spurious counterexample $x^*$ as a precise witness to the abstraction error. By performing a directional analysis at $x^*$, we transform a failed proof into a targeted guide for the branching process.

\section{Conclusion}
In this paper, we presented DRG-BaB, a complete verification framework that bridges the gap between dataflow abstraction and the CEGAR paradigm. By reformulating the BaB process as a CEGAR loop, we shifted the branching strategy from generic uncertainty reduction to a goal-driven refinement process guided by spurious counterexamples. Our analysis revealed that the spurious counterexample serves as a critical witness, pinpointing exactly where the convex relaxation is too loose relative to the safety specification. Furthermore, we validated the "Orthogonality Hypothesis," showing that branching resources are most effectively utilized when focusing on static relaxation errors that are orthogonal to the dynamic lower-bound optimization performed by the solver.

Extensive experiments on standard benchmarks confirmed that this CEGAR approach significantly prunes the search tree compared to heuristics based on interval width or average sensitivity. These findings suggest that tighter integration between the diagnostic outputs of the abstraction layer and the discrete branching logic is essential for scaling complete verification to deeper and more complex neural architectures. Future work will explore applying this directional analysis to more complex non-linearities, such as Sigmoid or Attention mechanisms in Transformer architectures.

\bibliographystyle{splncs04}
\bibliography{mybibliography}

\end{document}